\documentclass[12pt,twoside]{article}
\input{epsf.sty}

\def\mydate{21 September 1999}
\def\ignore#1{{}}

\newcommand{\beeq}{\begin{equation}}
\newcommand{\eneq}{\end{equation}}
\newcommand{\beqn}{\begin{eqnarray}}
\newcommand{\eeqn}{\end{eqnarray}}

\def\la{\raise.16ex\hbox{$\langle$} \, }
\def\ra{\, \raise.16ex\hbox{$\rangle$} }
\def\ran{\raise.16ex\hbox{$\rangle$} }
\def\go{\rightarrow}

\def\psibar{ \psi \kern-.65em\raise.6em\hbox{$-$}\lower.6em\hbox{} }
\def\Psibar{ \Psi \kern-.77em\raise.6em\hbox{$-$} }
\def\Phibar{ \Phi \kern-.77em\raise.6em\hbox{$-$} }

\begin{document}

\thispagestyle{empty}

\baselineskip=12pt

{\small \noindent Revised. \mydate \hfill To appear in Phys Rev Lett}

%

\baselineskip=40pt plus 1pt minus 1pt

\vskip 3cm

\begin{center}

\ignore{
{\LARGE\bf {Stable monopole and dyon solutions }}\\
{\LARGE\bf {in the Einstein-Yang-Mills theory}}\\
{\LARGE\bf {in asymptotically  anti-de Sitter space}}\\
}

{\Large\bf {Stable monopole and dyon solutions in the Einstein-Yang-Mills
theory in asymptotically anti-de Sitter Space}}\\

\vspace{1.5cm}
\baselineskip=20pt plus 1pt minus 1pt

{\large   Jefferson Bjoraker and Yutaka Hosotani}\\
\vspace{.1cm}
{\it School of Physics and Astronomy, University of Minnesota}\\  
{\it  Minneapolis, MN 55455, U.S.A.}\\ 
\end{center}

\vskip 3.cm
\baselineskip=20pt plus 1pt minus 1pt

\begin{abstract}
A continuum of new monopole and dyon solutions in the Einstein-Yang-Mills
theory in asymptotically anti-de Sitter space are found.  They are
regular everywhere and specified with their   mass, and non-Abelian
electric and magnetic charges. 
A class of monopole solutions which have no node in non-Abelian magnetic
fields are shown to be stable against spherically symmetric linear
perturbations.
\end{abstract}

\newpage

\vskip 1cm

Soliton and black hole solutions to the Einstein-Yang-Mills (EYM)
equations have generated considerable interest this past decade
\cite{BARTNIK}-\cite{VOLKOV2}. In flat space there can be no static
soliton solution in the pure Yang-Mills theory \cite{Deser1}.  Even
unstable static solutions cannot exist.
The presence of gravity provides attractive force which can bind
non-vanishing Yang-Mills fields together.  Such static 
 solutions to the EYM equations  have been  found  in
asymptotically Minkowski  or  de Sitter space.  They are all
purely magnetic. 
\ignore{The EYM equations were solved with  
the t' Hooft-Polyakov ansatz \cite{THOOFT} for the gauge fields which
excludes the electric components. This is justified}
In asymptotically Minkowski space-time the electric components are
forbidden in static solutions \cite{GALTSOV}.  All of these EYM solitons
and black holes were shown to be unstable, where the number of unstable
modes is equal to twice the number of times the magnetic component of the
gauge field crosses the axis \cite{VOLKOV3}.

If the spacetime is modified to include the cosmological constant,
the no-go theorems \cite{GALTSOV} forbidding the electric components of
the gauge fields fail, thus, permitting dyon
solutions. We have found that in asymptotically de
Sitter spacetime, a non-zero
electric component to the gauge fields causes $\sqrt{-g}$ to diverge at
the cosmological horizon, thus excluding  dyon solutions.

In this letter, we present new monopole and dyon solutions in
asymptotically anti-de Sitter (AdS) space.  They are regular everywhere.
In asymptotically AdS space there are solutions with no 
cosmological horizon and dyons solutions are allowed. Furthermore,
we have found a continuum of solutions where the gauge fields have no
nodes, corresponding to  stable monopole and dyon
solutions. 

AdS space has attracted huge interest recently.
The BTZ black holes in three-dimensional AdS space provide valuable
information about black hole thermodynamics and quantum
gravity\cite{BTZ}.  In four dimensions  linearly stable
black hole solutions have been found in asymptotically AdS
space\cite{WINSTANLEY}.   The correspondence between four-dimensional
super-conformal field theory and type IIB string theory on AdS$_5$ has
been  established\cite{ads-cft}.  In this Letter we shall show that
even in a  simple Einstein-Yang-Mills system stable monopole and dyon
solutions  exist in four-dimensional asymptotically AdS space, which we
believe leads to further understanding of rich structure of quantum
field theory in AdS space as well as profound implications to the
evolution of  the early universe.

We start with  the most  general 
expression for the spherically symmetric SU(2) gauge field 
in the singular gauge \cite{WITTEN}:
\begin{equation}
A=\frac{1}{2e} \, \Big\{ u\tau_3dt + \nu \tau_3 dr
+ (w\tau_1+\tilde{w}\tau_2)d\theta
+(\cot\theta\tau_3+w\tau_2-\tilde{w}\tau_1)\sin\theta d\phi\Big\} ,
\label{FIRST_A}
\end{equation}
where $u$, $\nu$, $w$ and $\tilde{w}$ depend on $r$ and $t$. 
The regularity at the origin imposes the boundary conditions
$u=\nu=0$ and $w^2 + \tilde w^2 = 1$ at $r=0$.  Under a residual $U(1)$
gauge transformation $S = e^{i \tau_3 z(t,r)/2}$, 
$(u, \nu) \go (u+ \dot z, \nu + z')$ and 
$w+ i\tilde w \go e^{iz} (w+ i\tilde w)$.  
The spherically symmetric metric is written as 
\begin{equation}
ds^2 = -\frac{H dt^2}{p^2} +\frac{dr^2}{H} 
+R^2 (d\theta^2 + \sin^2\theta d\phi^2),
\label{GMUNU}
\end{equation}
where $H$, $p$, and $R$ are functions of $t$ and $r$, in general.  

We first look for static regular soliton solutions in the $\nu=0$
gauge.   Since the Yang-Mills equations imply $\tilde{w} = Cw$ where $C$
is some constant, another gauge transformation allows us to set 
$C = 0$, or $\tilde{w} = 0$.  Without loss of generality one can choose
$w(0)=1$.
$u$ can be interpreted as the electric part of the gauge fields
and $w$ as the magnetic part.    The Schwarzshild gauge $R=r$ is taken
for the metric. The coupled static EYM equations of motion are 
\beqn
\left(\frac{H}{p}w^{\prime}\right)^{\prime} 
&=& -\frac{p}{H}u^2w-\frac{w}{p}\frac{(1-w^2)}{r^2} 
\label{YM1}  \\
\left(r^2pu^{\prime}\right)^{\prime}& =& \frac{2p}{H}w^2u  
\label{YM2}\\
p^{\prime} &=&
-\frac{2v}{r}p\left[(w^{\prime})^2+\frac{u^2w^2p^2}{H^2}\right] 
\label{Ein1}  \\
m' &=& v\left[
\frac{(w^2-1)^2}{2r^2}+\frac{1}{2}r^2p^2(u^{\prime})^2
  +H(w^{\prime})^2+\frac{u^2w^2p^2}{H} \right] ~.
\label{Ein2}
\eeqn
Here $H(r) = 1 - {2m(r)}/{r} - \Lambda r^2/3$ and $v ={G}/{4\pi e^2}$.
$\Lambda$ is the cosmological constant.  We are looking for a solution
which is regular everywhere and has finite ADM mass $m(\infty)$.

The presence of the cosmological constant affects the behavior of the 
soliton solutions significantly.  For $\Lambda=0$,  both $H$ and $p$
approach a constant value as $r \go \infty$. Consequently $w^2 \go 1$ from
(\ref{YM1}) and (\ref{Ein2}).   For $\Lambda \not= 0$, $w$ need not 
have an asymptotic value $\pm 1$.  It is known that Eq.\ (\ref{YM2})
implies the absence of  solutions with $u(r)\not= 0$ in the 
$\Lambda=0$ case \cite{GALTSOV}.   A similar argument applies to the $\Lambda>0$ case
in which there appears a cosmological horizon at $r=r_h$; $H(r_h)=0$.
At the horizon either $u$ or $w$ must vanish to have regular solutions
to Eqs.\ (\ref{YM1})-(\ref{Ein2}).   Eq.\ (\ref{YM2}) implies that 
$u(r)$ is either identically zero or a monotonic function of $r$ for 
$r< r_h$. Hence, $w(r_h)=0$ and $u(r_h) \not= 0$ if $u'(0) \not= 0$. 
However, Eq.\ (\ref{YM1}) implies 
\begin{equation}
 \frac{Hw^{\prime}}{pw} \Bigg|^{r_h}_{r_1} 
=  -\int_{r_1}^{r_h} dr \Bigg\{
\frac{H}{p}\left(\frac{w^{\prime}}{w}\right)^2
  +\frac{p u^2}{H}
+\frac{(1-w^2)}{pr^2} \Bigg\} 
\label{sum_rule1}
\end{equation}
provided $w(r) >0$ for $r_1 \le  r<r_h$.  The r.h.s.\ of
(\ref{sum_rule1}) diverges whereas the l.h.s.\  remains finite
if $w(r_h)=0$ and $u(r_h)\not= 0$.  This
establishes that $u(r)=0$ in the  de Sitter space.  All solutions
are purely magnetic.  Suppose that $0< w^2 \le 1$ for $0 \le r \le r_h$
and take $r_1=0$ in (\ref{sum_rule1}). As $w'(0)=0$, 
 the l.h.s.\ vanishes whereas the r.h.s.\ is negative definite.
This implies that $w$ satisfying $w^2 \le 1$ for $r \le r_h$ must vanish
somewhere between 0 and $r_h$.  It has been known that in all solutions
in asymptotically de Sitter space $w(r)$ has at least one node.

The situation is quite different in  asymptotically AdS space
($\Lambda<0$).  $H(r)$ is positive everywhere.  All equations are
consistently solved if $u(r) \sim u_0 + (u_1/r) + \cdots$ and 
 $w(r) \sim w_0 + (w_1/r) + \cdots$ for large $r$.  There is no
restriction on the value of $u_0$ or $w_0$.  Furthermore $w(r)$ can
be nodeless.

Solutions are classified by the ADM mass, $M=m(\infty)$, electric and
magnetic charges, $Q_E$ and $Q_M$.  From the Gauss flux theorem 
\beeq
\pmatrix{Q_E\cr Q_M\cr}
 = {e\over 4\pi} \int dS_k \, \sqrt{-g} \, 
\pmatrix{ F^{k0}\cr \tilde F^{k0} \cr}
\label{charge1}
\eneq
are conserved, but are also gauge-dependent.  With the ansatz  in the
singular gauge
\beeq
\pmatrix{Q_E\cr Q_M\cr}
 = \pmatrix{u_1 p_0\cr 1 - w_0^2\cr} {\tau_3\over 2}
\label{charge2}
\eneq
where $p(r) = p_0 + (p_1/r) + \cdots$ etc.  If $(u,w,m,p)$ is a
solution, then $(-u,w,m,p)$ is also a solution.  Dyon solutions come in
a pair with $(\pm Q_E, Q_M, M)$.

The solutions to Eqs.\ (\ref{YM1}) to (\ref{Ein2}), for $\Lambda < 0$, are
evaluated numerically. The procedure is to solve these equations at $r=0$
in terms of two free adjustable  parameters $a$ and $b$ and
`shoot' for solutions with the desired asymptotical behavior. The
behavior of solutions  near the origin
are
\begin{eqnarray}
w(r) &=& 1-br^2 \cr
m(r) &=& v(2b^2+\frac{1}{2}a^2)r^3 \cr
p(r) &=& 1 - (4b^2 + a^2)vr^2 \cr
u(r) &=& ar+\frac{a}{5}(-2b +\frac{1}{3}\Lambda + 2v(a^2+4b^2))r^3 ~.
\label{origin}
\end{eqnarray}

Purely magnetic solutions (monopoles) are found by setting $a=0$ 
($u=0$). Varying the initial condition parameter $b$,  a continuum of
monopole solutions were obtained, which are   similar to the black hole
solutions found in ref. \cite{WINSTANLEY}, but are regular in the entire
space.  $w$ crosses the axis an arbitrary number of times depending on
the value of the adjustable shooting parameter $b$.  Typical solutions
are displayed  in fig.\ 1.

\begin{figure}[th]\centering
 \leavevmode 
\vskip -.5cm
\mbox{
\epsfxsize=9.0cm \epsfbox{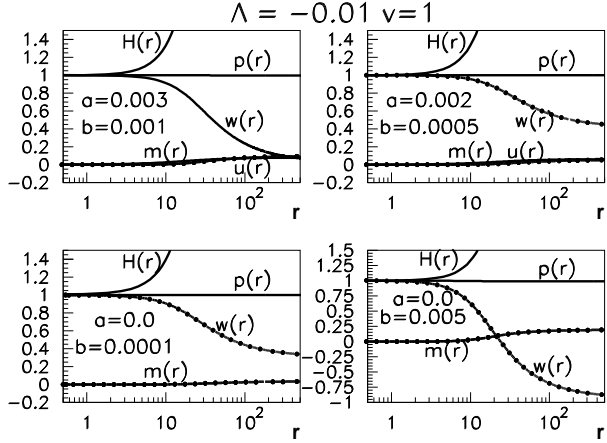}}
\caption{Monopole and dyon solutions for  $\Lambda = -0.01$ and $v=1$.
(a) Monopoles: $(a,b)=(0, 0.001)$ and $(0, 0.005)$.  $(w,m)$ at
$r=\infty$ are $(0.339, 0.034)$ and $(-0.878, 0.191)$, respectively.
(b) Dyons:  $(a,b)=(0.003, 0.001)$ and $(0.002, 0.0005)$.
$(w,u, m)$ at $r=\infty$ are $(0.031, 0.080, 0.099)$ and 
$(0.421,0.064,0.056)$, respectively.}
\label{fig1}
\end{figure}

The behavior of $m$ and $p$ is similar to that of the asymptotically de
Sitter solutions  previously considered \cite{VOLKOV}. In contrast, as
shown in fig.\ \ref{fig1}, there exist solutions where $w$ has no
nodes. These solutions are of particular interest because they are shown
to be stable against linear perturbations.

If the adjustable shooting parameter $a$ is chosen to be non-zero,  we
find dyon solutions.
As shown in fig.\ \ref{fig1}, the electric component, $u$, of the YM
fields starts at zero and monotonically increases to some finite
value. The behavior of $w$, $m$ , $H$, and $p$ is similar to that in the
monopole solutions.

Again we find a continuum of solutions where $w$ crosses the axis an
arbitrary number of times depending on $a$ and $b$. Similarly there
exist solutions where $w$ does not cross the axis. 

Solutions are found for a continuous set of the parameters $a$ and
$b$.  This is in sharp contrast to the $\Lambda \ge 0$ case, in which
only a discrete set of solutions are found.   For some values of $a$ and
$b$,  solutions blow up, or  the function $H(r)$ crosses the
axis and becomes negative. 
One example of solutions near the critical value
[$(a,b)=(0.01, 0.69)$] is displayed in fig.\ 2.  $H(r)$ becomes very
close to zero at $r \sim 1$. It has $(Q_E, Q_M, M) \sim
(0.015,0.998,0.995)$.

\begin{figure}[tbh]\centering
\leavevmode 
\mbox{
\epsfxsize=8.cm \epsfbox{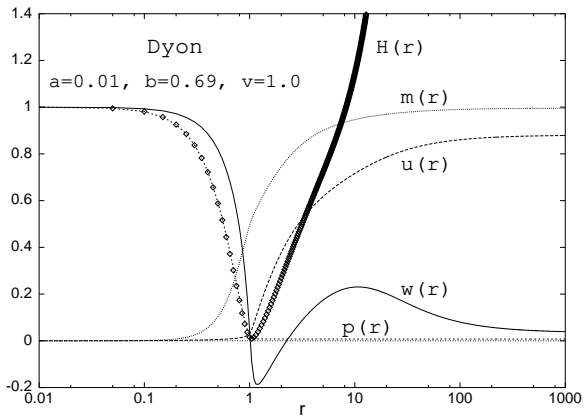}}
\caption{Dyon solution for  $\Lambda = -0.01$, $v=1$, $a=0.01$ and 
$b=0.69$.  $b$ is close to the critical value $b_c=0.7$.  $H$ almost
hits the axis around $r=1$.}
\label{fig2}
\end{figure}

\begin{figure}[tbh]\centering
\leavevmode 
\mbox{
\epsfxsize=8.cm \epsfbox{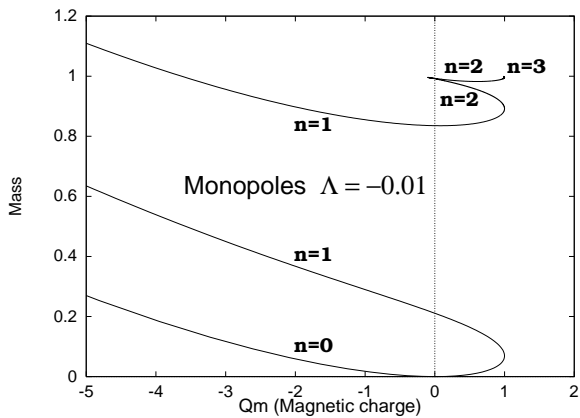}}
\caption{Mass $M$ is plotted as a function of magnetic charge $Q_M$
for monopole solutions.  The number of nodes, $n$,  in $w(r)$ is also 
marked.}
\label{fig3}
\end{figure}

In fig.\ 3 $M$ is plotted as a function of $Q_M$ for monopole
solutions. The behavior of the solutions near $b=0.7$ needs more careful
analysis, although we did find that when  $b > 0.7$, all the solutions
appeared to have a horizon.

We have found that  dyon solutions cover a good portion of the
$Q_E$-$Q_M$ plane.  There are solutions with $Q_M=0$ but
$Q_E\not=0$, which has non-vanishing $w(r)$. In the shooting parameter
space $(a,b)$,  these solutions correspond not exactly, but almost to a
universal value for $b \sim 0.0054$.  See fig.\ 4.  We have not
understood why it should be so.

\begin{figure}[th]\centering
 \leavevmode 
\mbox{
\epsfxsize=8.cm \epsfbox{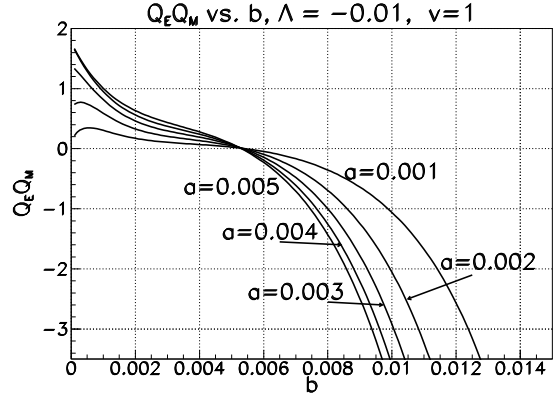}}
\caption{$Q_E \cdot Q_M$ is plotted as a function of $b$  with $a$
fixed.  $Q_M$ vanishes around $b=0.054$.}
\end{figure}

It has been shown that the BK solutions and the de Sitter-EYM
solutions are unstable
\cite{ZHOU}\cite{GREENE}\cite{VOLKOV2}\cite{BRODBECK}. In
contrast,  the AdS black hole solutions were shown to be
stable \cite{WINSTANLEY} for $u=0$.  We shall show that the monopole
solutions without nodes presented in this paper are stable against
spherically symmetric linear fluctuations.

In examining time-dependent fluctuations around monopole solutions it is
convenient to work in the $u(t,r)=0$ gauge.  Solutions have
non-vanishing $w(r)$, $p(r)$, and $H(r)$, but $\tilde w(r) = \nu(r)
=0$.     Linearized equations for $\delta w(t,r)$, $\delta \tilde
w(t,r)$, $\delta \nu(t,r)$, $\delta p(t,r)$, and $\delta H(t,r)$
have been derived in the literature \cite{VOLKOV2}.   Fluctuations decouple
in two groups.   $\delta w(t,r)$,   $\delta p(t,r)$, and $\delta H(t,r)$
form even-parity perturbations, whereas $\delta \tilde w(t,r)$ and 
$\delta \nu(t,r)$ form odd-parity perturbations.  The linearized 
equations imply that  $\delta p(t,r)$, and $\delta H(t,r)$ are
determined by $\delta w(t,r)$, and  $\delta \tilde w(t,r)$
 by $\delta \nu(t,r)$.

$\beta(t,r) = {r^2 p \delta\nu / w}= e^{-i\omega t}\beta(r)$
satisfies 
$\big\{ - (d/d\rho)^2 + U_\beta(\rho) \big\}  \beta
= \omega^2 \beta$ where 
\beeq
U_\beta = \frac{H}{r^2p^2}(1+w^2)
+\frac{2}{w^2}\left(\frac{dw}{d\rho}\right)^2 
\label{U_nu}
\eneq
and ${d\rho/ dr} = {p/ H}$.
The range of $\rho$ is finite: $0 \le \rho \le  \rho_{\rm max}$. 
Eq.\ (\ref{U_nu}) is of the form of the Schr\"odinger equation
on a one-dimensional interval.  When $w(r)$ in the monopole solution
has no node ($w>0$ for all $r$), then $U_\beta >0$ is a smooth potential
so that $\beta$ is a smooth function of $\rho$ in the entire range
satisfying $\beta \beta'=0$ at $\rho=0$ and $\rho_{\rm max}$.  The 
eigenvalue $\omega^2$ is positive-definite, i.e.\ the solution is stable
against odd-parity perturbations.  If $w(r)$ has $n$ nodes, i.e.
$w(r_j)=0$ ($j=1, \cdots, n$), the potential $U_\beta$ develops
$(\rho-\rho_j)^{-2}$ singularities.   The solution $\beta(\rho)$
to (\ref{U_nu}) is no longer regular at $\rho=\rho_j$ so that the 
positivity of the differential operator $-d^2/d\rho^2$ is not
guaranteed.  Indeed, Volkov et al.\ have proven for the 
BK solutions that there appear exactly $n$ negative eigenmodes
($\omega^2<0$) if $w$ has $n$ nodes \cite{VOLKOV3}.  Their argument
applies to our case without modification.  One concludes that the
solutions with nodes in $w$ are unstable against parity-odd perturbations.

Similarly  parity-even perturbations $\delta w(t,r) = e^{-i\omega t}
\delta w(r)$ satisfies the Schr\"odinger equation with a potential
\beeq
U_w = \frac{H (3w^2-1)}{p^2r^2} 
+ 4 v \frac{d}{d\rho}\left(\frac{Hw'^2}{pr}\right) ~.
\label{U_w}
\eneq
$U_w(\rho)$ is not positive definite, but is regular
in the entire range $0 \le \rho \le \rho_{\rm max}$.   We have solved
the  Schr\"odinger equation for $\delta w$ numerically for typical
monopole solutions.  
The potential is displayed in fig.\ 5 for the
solutions with $(a,b)=(0, 0.001)$ and $(0, 0.005)$.  The former has no
 node in $w$, while the latter has one node.  The asymptotic value
$w(\infty)$  is 0.339 or $-0.878$, respectively. The lowest eigenvalue
$\omega^2$ is found to be 0.028 or 0.023.  Hence, these solutions are
stable against  parity-even perturbations.  Note that in the $\Lambda<0$
case some of the $n=1$ solutions are stable against parity-even
perturbations, while in the $\Lambda \ge 0$ case they are unstable.

\begin{figure}[tbh]\centering
 \leavevmode 
\mbox{
\epsfxsize=8.cm \epsfbox{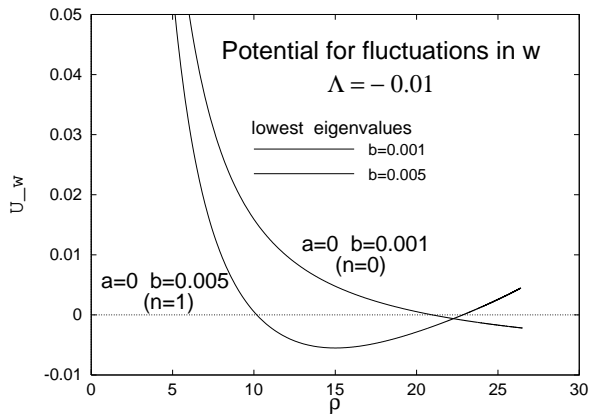}}
\caption{The potential $U_w(\rho)$ in (12) for the monopole solutions
with $b=0.001$ and 0.005.  The number of nodes ($n$) in $w(r)$
is  0 and 1, respectively.  The lowest eigenvalue
$\omega^2$ is found to be 0.028 or 0.023, respectively, implying the
stability of the solutions.}
\end{figure}

\ignore{
For $(a,b)=(0, 0.001)$, $w(\infty)=0.339$.
Monotonically decreasing $U_w(\rho)$ becomes negative
at $\rho=21$ (?) and reach $-0.002$ (?) at $\rho_{\rm max}$.
The lowest eigenvalue $\omega^2$ is 0.028.
For $(a,b)=(0, 0.005)$, $w(\infty)=-0.878$.
$U_w(\rho)$ takes a minimum value $-0.005$ (?) at $\rho=14.?$.
The lowest eigenvalue $\omega^2$ is 0.023.
Hence, these solutions are
stable against  parity-even perturbations.  Note that in the $\Lambda<0$
case some of the $n=1$ solutions are stable against parity-even
perturbations, while in the $\Lambda \ge 0$ case they are unstable.
}

In the present paper a continuum of new  monopole and dyon solutions to
the  EYM equations in asymptotically AdS space have been found. There
are solutions with no node in the magnetic component $w(r)$ of the
$SU(2)$ gauge fields. The monopole solutions with no node in $w$ have
been shown to be stable against spherically symmetric perturbations.
The  stability of those solutions with non-zero electric fields is
currently under investigation.   As the monopole and dyon solutions are
found in a continuum set, the dyon solutions without nodes in $w(r)$
are also  expected to be stable.

The existence of stable monopole and dyon configurations may have
tremendous consequences in cosmology if the early universe ever
was in the AdS phase.  Stable solutions exist only 
with a negative cosmological constant.  The existence of the
boundary in the AdS space must be playing a crucial role.
The connection to the AdS/CFT correspondence\cite{ads-cft} is yet to be
explored.   A more thorough analysis of the solutions with varying
$\Lambda$ as well as blackhole solutions will be presented  in separate
publications.

\vskip 1cm

\leftline{\bf Acknowledgments}

This work was supported in part    by the U.S.\ Department of
Energy under contracts DE-FG02-94ER-40823 and DE-FG02-87ER40328.


\vskip 1cm 

\begin{thebibliography}{99}
\parskip=0pt
\small
\bibitem{BARTNIK}
         R. Bartnik and J. McKinnon,
         Phys. Rev. Lett {\bf 61 }, 141 (1988).
\bibitem{ZHOU}
         N. Straumann and Z. Zhou,
         Phys. Lett. {\bf B237}, 353 (1990);
         Z. Zhou and N. Straumann,
         Nucl. Phys. {\bf B360}, 180 (1991).
\bibitem{BIZON} 
         P. Bizon, 
         Phys. Rev. Lett. {\bf 64 \rm}, 2844 (1990).
\bibitem{GREENE}
         B. Greene, S. Mathur, and C. O'Niell,
         Phys. Rev. D {\bf 47 \rm} 2242 (1993);
         K. Lee, V. Nair and E. Weinberg,
         Phys. Rev. Lett. {\bf 68 \rm}, 1100 (1992).
\bibitem{VOLKOV}
        M.S. Volkov, N. Straumann, G. Lavrelashvili, M. Huesler 
        and O. Brodbeck,
        Phys. Rev. D {\bf 54} 7243 (1996);
        T. Torii, K. Maeda and T. Tachizawa,
        Phys. Rev. D {\bf 52} R4272 (1995).
\bibitem{VOLKOV2}
        M. S. Volkov and D. Gal'tsov,
        hep-th/9810070 (1998).
\bibitem{Deser1}
         S. Deser, Phys. Lett. {\bf 64B}, 463 (1976).
\ignore{
\bibitem{THOOFT}
        G. 't Hooft, 
        Nucl, Phys. {\bf B 79}, 276 (1974);
         A. Polyakov,
         JETP Lett. {\bf 20} 194 (1974).
}
\bibitem{GALTSOV} 
         A. Ershov and D. Galt'sov, 
         Phys. Lett. A {\bf 138 \rm}, 160 (1989);
         Phys. Lett. A {\bf 150 \rm}, 159 (1990).
\bibitem{VOLKOV3}
        M.S. Volkov, O. Brodbeck,  G. Lavrelashvili, and N. Straumann, 
        Phys. Lett. {\bf B349}, 438 (1995). 

\bibitem{BTZ}
M. Banados, C. Teitelboim, and J. Zanelli, 
   Phys. Rev. Lett. {\bf 69} 1849 (1992); 
M. Banados, M. Henneaux, C. Teitelboim, and J. Zanelli, 
   Phys. Rev. {\bf D48} 1506 (1993); 



\bibitem{WINSTANLEY}
         E. Winstanley, Class. Quant. Grav. {\bf 16} 1963 (1999).
\bibitem{ads-cft}
J.\ Maldacena, Adv.\ Theoret.\ Math.\ Phys.\ {\bf 2}, 231 (1998);
E. Witten,  Adv.\ Theoret.\ Math.\ Phys.\ {\bf 2}, 253 (1998);
{\it ibid.} {\bf 2}, 505 (1998).
\bibitem{WITTEN}
         E. Witten,
         Phys. Rev. Lett. {\bf 38 \rm} 121 (1977).

\bibitem{BRODBECK}
        O. Brodbeck, M. Huesler, G. Lavrelashvili, N. Straumann 
        and  M.S. Volkov,
        Phys. Rev. D {\bf 54} 7338 (1996).        
\end{thebibliography}
\end{document}